\documentclass{cs19proc}

\usepackage{aas_symbols}
\usepackage[labelformat=simple]{subcaption}

\usepackage[breaklinks,colorlinks,citecolor=blue,urlcolor=blue]{hyperref}
\newcommand{\rot}{{\rm rot}}
\newcommand{\cyc}{{\rm cyc}}
\newcommand{\var}{{\rm var}}

\newcommand{\II}{{\sc \romannumeral 2}}
\newcommand{\Teff}{\ensuremath T_{\rm eff}}

\editors{G.~A. Feiden}
\publisher{Zenodo}
\conference{The 19th Cambridge Workshop on Cool Stars, Stellar Systems, and the Sun}
\conferencedate{2016}

\title{Dynamo Sensitivity in Solar Analogs with 50 Years of Ca II H \& K Activity}
\author{Ricky Egeland$^{1,2}$, Willie Soon$^{3}$, Sallie Baliunas$^{4}$, Jeffrey C. Hall$^{5}$, Alexei A. Pevtsov$^{6,7}$, Gregory W. Henry$^{8}$}

\affiliation{$^{1}$ High Altitude Observatory, Boulder, Colorado, USA \\
             $^{2}$ Montana State University, Bozeman, Montana, USA \\
             $^{3}$ Harvard-Smithsonian Center for Astrophysics, Cambridge, Massachusets, USA \\
             $^{4}$ No Affiliation \\
             $^{5}$ Lowell Observatory, Flagstaff, Arizona, USA \\
             $^{6}$ Natioanl Solar Observatory, Sunspot, New Mexico, USA \\
             $^{7}$ ReSoLVE Centre of Excellence, Space Climate Research Unit, University of Oulu, Finland \\
             $^{8}$ Center of Excellence in Information Systems, Tennessee State University, Nashville, Tennessee, USA
}

\shorttitle{Solar Analogs with 50 Years of Ca II H \& K Activity}
\shortauthors{Ricky Egeland}

\abs{The Sun has a steady 11-year cycle in magnetic activity most
  well-known by the rising and falling in the occurrence of dark
  sunspots on the solar disk in visible bandpasses.  The 11-year cycle
  is also manifest in the variations of emission in the Ca II H \& K
  line cores, due to non-thermal (i.e. magnetic) heating in the lower
  chromosphere.  The large variation in Ca II H \& K emission allows
  for study of the patterns of long-term variability in other stars
  thanks to synoptic monitoring with the Mount Wilson Observatory HK
  photometers (1966-2003) and Lowell Observatory Solar-Stellar
  Spectrograph (1994-present).  Overlapping measurements for a set of
  27 nearby solar-analog (spectral types G0-G5) stars were used to
  calibrate the two instruments and construct time series of magnetic
  activity up to 50 years in length.  Precise properties of
  fundamental importance to the dynamo are available from Hipparcos,
  the Geneva-Copenhagen Survey, and CHARA interferometry.  Using these
  long time series and measurements of fundamental properties, we do a
  comparative study of stellar ``twins'' to explore the sensitivity of
  the stellar dynamo to small changes to structure, rotation, and
  composition.  We also compare this sample to the Sun and find hints
  that the regular periodic variability of the solar cycle may be rare
  among its nearest neighbors in parameter space.
}

\begin{document}

\maketitle

%%%
%%%
%%%
\section{Introduction}

Emission in the Ca \II{} H \& K line cores has long been known to be a
good proxy for magnetic activity in the Sun \citep{Hall:2008}.
\cite{Wilson:1978} was the first to use this emission to demonstrate
the magnetic variability for an ensemble of Sun-like stars, using a
decade of synoptic Ca \II{} H \& K observations from the Mount Wilson
Observatory (MWO).  The MWO HK project began in 1966 and continued
until 2003, with the largest compendium of stellar activity for 111
stars with up to 25 years of observations appearing in
\cite{Baliunas:1995}.  The MWO HK project was the basis of numerous
investigations of activity, its relationship to stellar age and
rotation, and implications for dynamo theory \citep[see][, and
  references therein]{Baliunas:1998}.A complimentary synoptic
observation program began at Lowell Observatory in the mid-1990's
using the Solar Stellar Spectrograph (SSS), designed to take low
resolution spectra covering the Ca \II{} H \& K region for the Sun and
stars with the same spectrograph \citep{Hall:1995,Hall:2007b}.  The
SSS program continues to this day, and 57 of its $\sim$100 targets
overlap with the MWO HK project.  We combine the data from these two
surveys making time series of nearly 50 years in length.  This was
done for the first time in \cite{Egeland:2015} for the young solar
analog HD 30495.  In that case, the long time series allowed for the identification
of three and a half stellar cycles, with a mean period of $\sim$12
years, for a star that previously appeared to be acyclically variable.
Work is ongoing to calibrate, combine, and analyze MWO+SSS time series
for a sample of 27 solar analog stars with $0.59 \leq (B-V) \leq 0.69$
\citep{CayreldeStrobel:1996}, in order to understand the solar dynamo
in the stellar context.  In particular, we seek to better understand
(1) whether the pattern of solar variability is common among Sun-like
stars (2) how the patterns of long-term variability in the ensemble
depend on stellar properties such as mass, luminosity, radius,
metalicity, and rotation.  Preliminary results from this project were
presented at this conference
\citep{Egeland:2016:cs19talk,Egeland:2016:cs19poster} and are
summarized in these proceedings.  The full details and final results
are to appear in a peer-reviewed journal in the near future.

%%%
%%%
%%%
\section{Solar-Analog Sample}

\begin{figure*}
  \centering
  \begin{subfigure}{0.95\columnwidth}
  \includegraphics[height=19em]{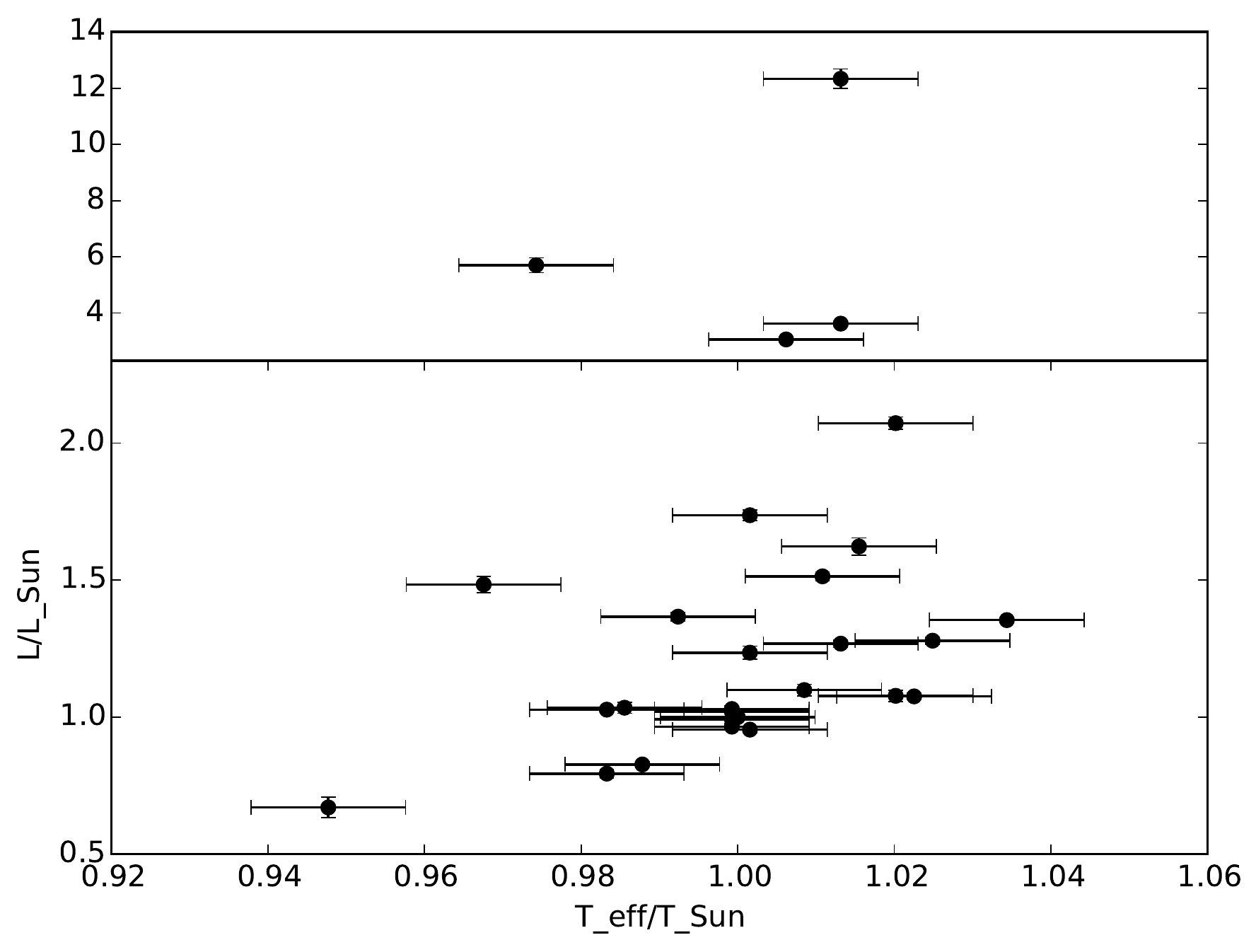}
  \caption{Luminosity vs. effective temperature}
  \label{fig:LvsT}
  \end{subfigure}\hfill
  \begin{subfigure}{0.95\columnwidth}
  \includegraphics[height=19em]{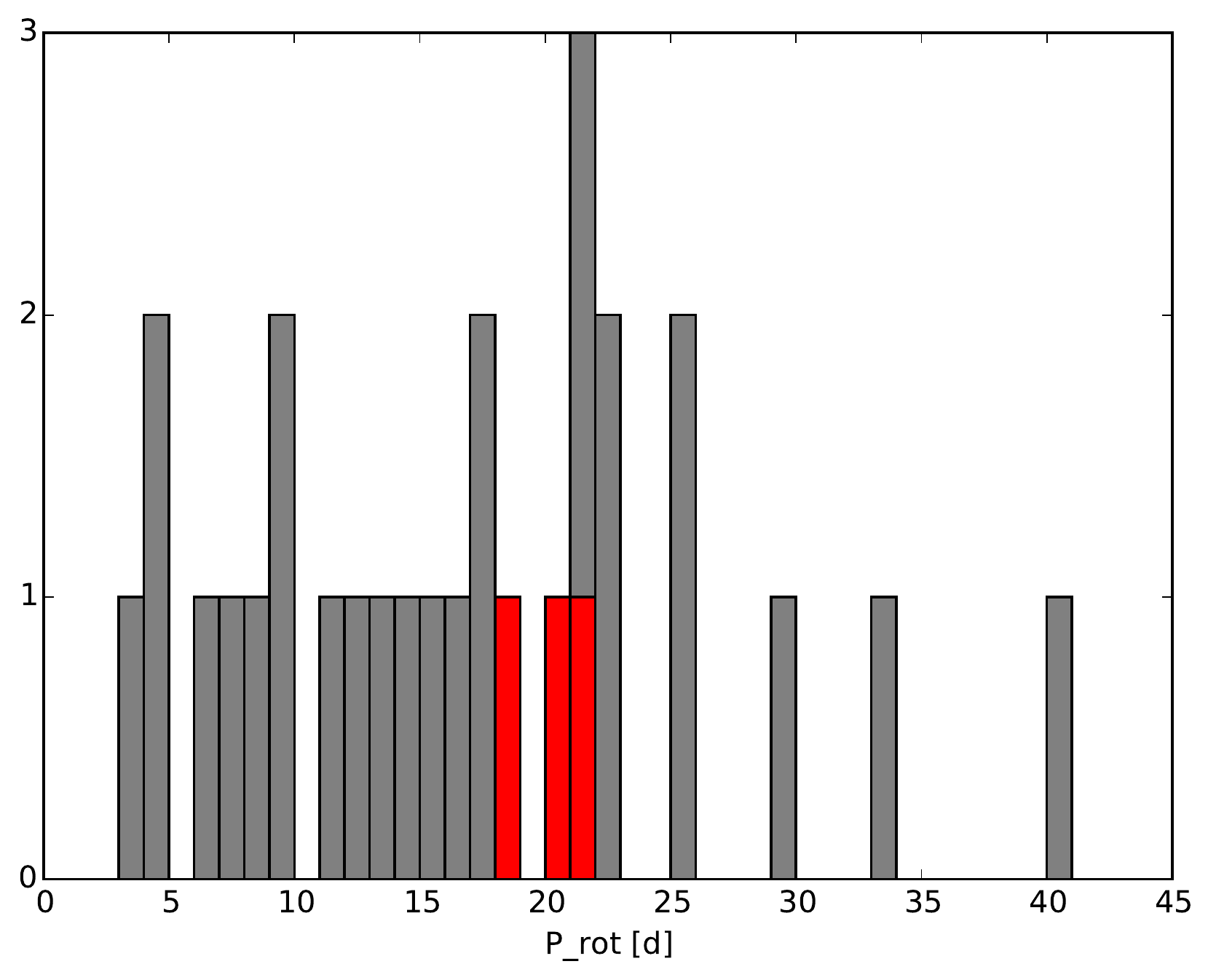}
  \caption{Rotation distribution}
  \label{fig:rot_dist}
  \end{subfigure}
  \caption{Stellar properties for the MWO+SSS solar analog sample.
    Note that the error bars in the luminosity are often hidden behind
    the data point.}
  \label{fig:props}
\end{figure*}

Our sample consists of the Sun and 27 solar-analog stars ($0.59 \leq
(B-V) \leq 0.69$) with synoptic observations from both MWO and the
SSS.  Of these, 20 stars (including the Sun) have activity time series
of nearly 50 years in length, with the remainder having somewhat less
coverage but nonetheless with at least 20 seasons of observations.
Our cut in $(B-V)$ color index keeps the sample within roughly 10\% of
the solar mass for stars firmly on the main sequence, though some
subgiants are in the sample as we shall see below.  From the
perspective of the stellar dynamo, perhaps a more important result of
this limitation is that the stars share a roughly similar luminosity,
which puts limits on the energy available to drive convection in our
sample.  Our limited parameter space is designed with the hope that a
larger fraction of the stars in the sample have dynamos driven by
processes similar to that of the Sun.  More massive stars with thin
convection zones and high convective energy, as well as less massive stars
with deep convection zones and lower convective energy are excluded
from this sample.

We characterize our sample using results from the Geneva Copenhagen
Survey (GCS) based on Str\"{o}mgren $ubvy\beta$ photometry and
Hipparcos parallaxes \citep{Holmberg:2009}.  Each star in our sample
is nearby and bright ($V < 7$), typically with a long literature of
observations.  From the GCS, we obtain the absolute magnitude,
effective temperature, and metalicity.  We convert the absolute
magnitudes from the GCS to luminosity using the bolometric correction
of \cite{Torres:2010}, based on the work of \cite{Flower:1996}.
Luminosity and effective temperature are then converted into stellar
radius using the Stephan-Boltzman law.  Figure \ref{fig:LvsT} shows
luminosity versus effective temperature for our sample.  Effective
temperatures are within 5\% of the solar nominal value of 5772 K, with
most stars within 2$\sigma$ of the solar temperature, where $\sigma =
57$ K is the estimated measurement uncertainty for GCS temperatures
\citep{Holmberg:2009}.  Five stars in our sample have $L > 2 \L_\Sun$
and are thus appreciably evolved.  Excluding these five, luminosities
range from 0.67 to 1.74 $L_\Sun$.  The median temperature, luminosity,
and radius for our sample is 1.00, 1.17, and 1.06 the solar value,
respectively.  Metalicities range from -0.78 to +1.3 dex with a median
value of -0.1 dex.

Rotation periods are taken from various literature sources, the
majority coming from the rotation studies of \cite{Donahue:1996} and
\cite{Baliunas:1996b}, who used a periodogram analysis on seasonal MWO
HK time series to measure rotation.  Figure \ref{fig:rot_dist} shows a
histogram of the rotation periods for our sample.  Rotation periods
for three stars, estimated from their projected rotation velocities
($v\sin(i)$) and radii, are shown by red bins
and are only a lower limit dependant on the inclination, $P_\rot /
\sin (i)$.  All other rotation periods are measured using a
periodogram analysis of $S$-index or Str\"{o}mgren $by$ photometry
time series, which are modulated by the passage of active regions on
the stellar surface.  Work is ongoing to measure rotation from such
time series for the three $P_\rot / \sin(i)$ stars.  Figure
\ref{fig:rot_dist} shows that our sample has relatively uniform
sampling in rotation up to about $P_\rot = 22$ days, after which the
sampling is sparse.  Six stars have a rotation within 20\% of the
solar rotation period, here taken to be 25 days, although the three
$P_\rot / \sin(i)$ stars may also have rotations within that range.
The median rotation period is 15 days.

In summary, our sample generally has properties close to solar
values, but the sample centroid is slightly more luminous and
metal-poor, and rotates faster than the Sun.

%%%
%%%
%%%
\section{Analysis Methods}

%%% Statistical Properties of S(t)
Consider the stellar dynamo to be an unknown function which maps
measurable global properties such as effective temperature,
luminosity, composition and rotation into a time varying, spatially
distributed magnetic field collapsed into a one-dimensional time
series by integrating chromospheric Ca \II{} HK emission over the
stellar surface.  Then with the dynamo \emph{inputs} characterized by
the GCS and rotation measurements described above, our next job is to
characterize the dynamo \emph{outputs} using our long $S$-index time
series.  Firstly, we characterize the statistical properties of the
variability using rank-based measures that are robust against outliers
and appropriate for use on non-Gaussian distributions.  For each
$S$-index time series, we calculate the median $\overline{S}$, the
upper $99^{\rm th}$ percentile and the lower $1^{\rm st}$ percentile.
The difference of these percentiles gives $A_{98}$, the amplitude of
the inner 98\% of the measurements.  This amplitude is designed to
estimate the total range of the measurements while being robust to
small numbers of outliers.  The $S$-index binned into 1-year observing
seasons and the amplitude $A_{98,s}$ is computed for each season.  We
report the median seasonal amplitude, $\overline{A}_{98,s}$ as an
estimate of the typical amplitude of variability in a 1-year period.
We thus obtain an estimate of the amplitude of long-term (decades) and
short-term (1 year) variability for each star.

%%% Period search in S(t)
We perform a Lomb-Scargle periodogram analysis on each of our
composite time series following the methods of \cite{Baliunas:1995}
and \cite{Horne:1986}.  We search for statistically significant peaks
by computing a power spectral density threshold above a false alarm
probability (FAP) of 0.1\%, the minimum confidence threshold for a
``poor'' cycle in \cite{Baliunas:1995}.  The FAP gives the probability
that a given periodogram peak is due to random noise, and the
confidence level that the signal exists is 1 - FAP > 99.9\%.  The top
three statistically significant peak periods, $P_\var$, are stored for
further consideration according to the quality metric, described
below.  In many previous works, \cite[e.g.][]{Baliunas:1995}, the top
two statistically significant periodogram peaks are reported as
``primary'' and ``secondary'' cycles.  One of our aims in this work is
to define a quantitative basis for classifying a periodogram peak as a
``cycle'', which satisfies our qualitative notions of what constitutes
a cycle.

%%% Description of Q_cyc
Using the solar cycle as the primary model of how we would like to
define a stellar activity cycle, we note two important qualities: (1)
the cycle pattern \emph{approximately} repeats for dozens of
iterations, lasting centuries (2) the cycle pattern is dominant; other
periodicities, if and when they are present, are of much lower
amplitude than the primary $\approx$11 year cycle.  We seek to define
a quality metric which can be used to find variations which have these
two characteristics.  By contrast the FAP of a
Lomb-Scargle periodogram peak at period $P_\var$, when
low, gives us confidence that a sinusoidal signal is \emph{present},
and not simply due to random noise.  Defining  ``noise'' to be
everything that is \emph{not} the primary long-term cycle in a record
of solar activity (i.e. rotational modulations; active region growth
and decay; other short-period variations that may be dynamo-related),
then we find that the solar cycle in MWO or SSS S-index has a
signal to noise ratio of $\approx$10.  Therefore in the search for
solar-like cycles, we are \emph{not} faced with the problem of
extracting a faint signal from noisy data.  FAP is therefore not an
appropriate tool to quantitatively compare stellar cycles.

Besides this, FAP scales with the number of data points in the time
series.  As a result, the 4-class system (poor, fair, good, excellent)
used in \cite{Baliunas:1995} cannot be applied to other data sets
which may have more or less observations.  For our nearly 50-year time
series, \emph{nearly every star has an ``excellent'' cycle}, even though
inspecting the time series one would have great difficulty finding the
purportedly ``excellent'' signal.

We therefore define a new quality metric:

\[  {\rm ASD} = \sqrt{\frac{2}{N} \, {\rm PSD}} \nonumber \]
\[  Q_\cyc = 100 \left(1 - 0.5 \frac{P_\var}{T} \right) \, {\rm ASD} \]

\noindent where ASD is the amplitude spectral density, and PSD is the
power spectral density, normalized by the variance of the data as
described in \cite{Horne:1986}, $T$ is the duration of the time
series, and $N$ is the number of samples.  With the PSD normalized by
the variance $\sigma^2$, ASD has units of $X \sigma_X^{-1} T^{-1}$,
where $X$ represents the units of the time series and $T$ represents
the time units.  For a pure sinusoidal signal of \emph{any} amplitude,
the ASD has a value of 1, indicating that the rms amplitude
of the signal is 1 $\sigma$.  ASD is therefore bounded from [0, 1].
The factor $(1 - 0.5 P_\var/T)$ is a penalty factor for infrequently
observed cycles.  If only one full cycle of a pure sinusoid is
observed, $Q_\cyc = 50$.  As $T \rightarrow \infty$ for a pure
sinusoid, there is no penalty and $Q_\cyc \rightarrow 100$.  $Q_\cyc$
is always positive so long as $P_\var > T$, which is ensured in our
analysis since we do not search for periods longer than the time
series.  Therefore, in general, $Q_\cyc$ ranges from [0, 100] with 100
only achievable with an infinite time series of a pure sinusoid.  The
solar cycle is not a pure sinusoid, and we do not have an infinite
record, so even in the best cases $Q_\cyc$ will be somewhat less than
100.  ASD is insensitive to the number of observations $N$, therefore
two separate instruments observing the same star during the same
period should in principle obtain the same $Q_\cyc$ even with
different sampling, which is not true of a quality scale based on FAP.

In an upcoming work, we will explore the properties and caveats
of $Q_\cyc$ in more detail, but so far we are satisfied with the
qualitative ranking of cycles by this metric.  For the Sun $Q_\cyc =
59$, and stars with $Q_\cyc > 50$ have cycles that are easy to
identify from simple inspection of the time series.  $Q_\cyc > 40$ are
still identifiable but not so obvious, and as $Q_\cyc
\rightarrow 0$ no obvious periodicity can be seen in the time series,
despite the FAP indicating that the $P_\var$ is statistically
significant.  The functional form and coefficient of 0.5 in the
observation time penalty factor are arbitrary, but they serve the
important purpose of reducing $Q_\cyc$ for relatively flat time series
that have a long-term trend and, therefore, a periodogram peak near the
window length.  Furthermore, our criteria that a ``cycle'' is
something that repeats warrants a penalty for any pattern that is only
seen once.

Finally, we evaluate the sensitivity of the stellar dynamo to
fundamental properties by examining pairs of stellar ``twins'' using
the Euclidean distance metric:

\[
  d(\mathbf{p}, \mathbf{q}) = \sqrt{ \sum_{i=1}^N (p_i - q_i)^2 }
\]

where $\mathbf{p}$ and $\mathbf{q}$ are stellar property vectors \{
$\Teff$, $R$, $P_\rot$ \} for two different stars, all measured in
solar units.  With this distance metric, stellar twins
are identified as those with a short distance.  We can then examine
the dynamo \emph{outputs} of stellar twins to answer the question:
\emph{Do identical stars have identical patterns of magnetic
  variability?}

Despite scaling to solar units, the relative importance
of these three parameters is not the same, because as can be seen in
Figure \ref{fig:props} the range of rotations
is much larger than the range of effective temperatures, for example.
However, it is reasonable to allow rotation to have more weight in the
distance metric than effective temperature, since rotation has a
larger effect on activity.

%%%
%%%
%%%
\section{Results}

\begin{figure*}[ht!]
  \includegraphics[width=\textwidth]{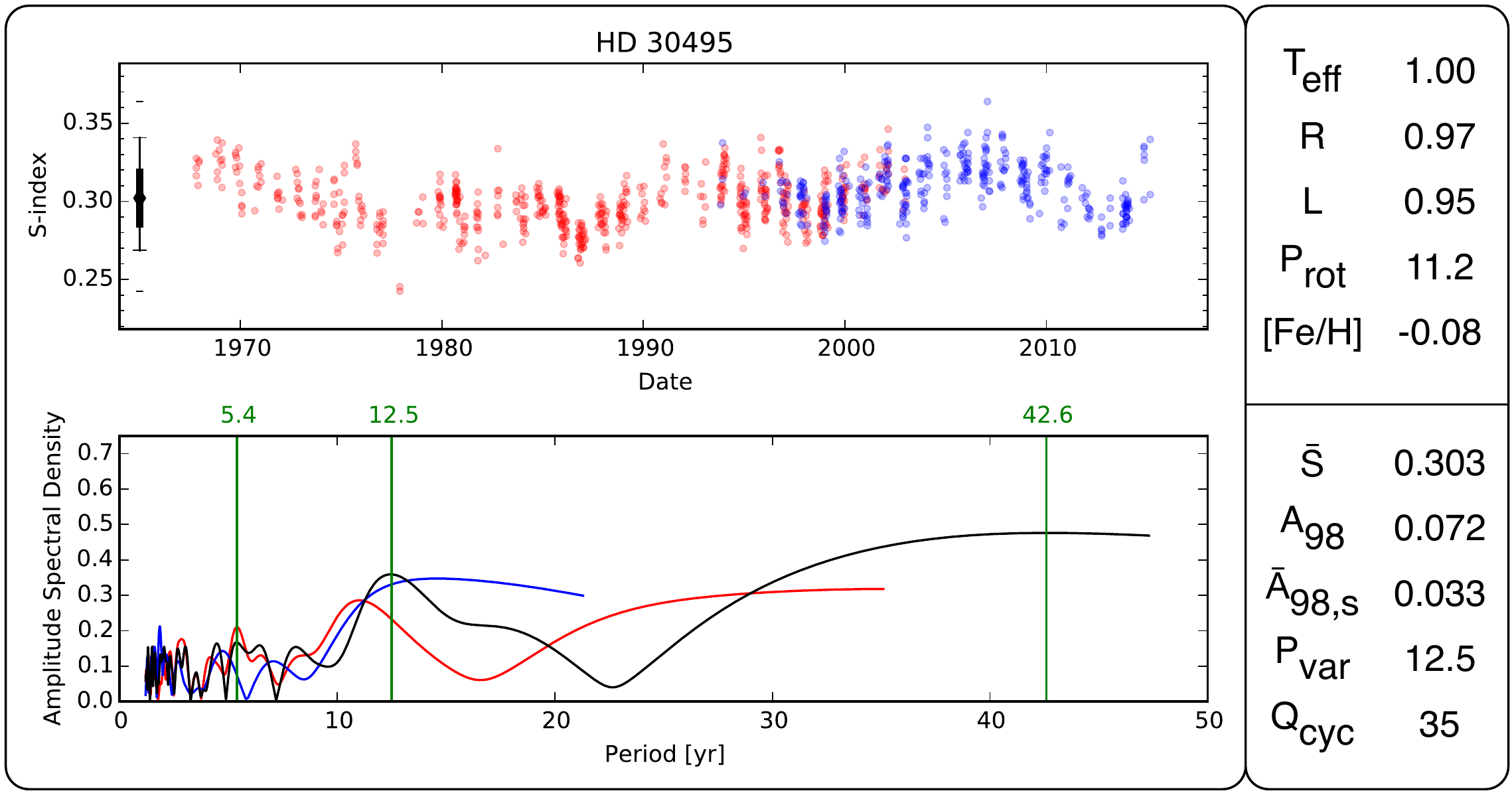}
  \caption{Time series, periodogram, stellar properties and variability
    measurements for HD 30495.}
  \label{fig:hd30495}
\end{figure*}

%%% Data is in the poster
Measurements for each star in our sample can be seen in our
poster\footnote{\url{https://doi.org/10.5281/zenodo.57921}}, which is
also published as part of these proceedings
\citep{Egeland:2016:cs19poster}.  A sample panel from the poster is
shown for HD 30495 in Figure \ref{fig:hd30495}, and the variability of
this star was studied in detail in \cite{Egeland:2015}.  The top plot
is the time series of MWO observations in red, and SSS observations in
blue.  The bottom plot is the periodogram for the time series shown as
an amplitude spectral density (ASD), with the black line utilizing all
observations, the red line utilizing only MWO and the blue line only
SSS.  The top table gives the stellar properties in solar units,
except for $P_\rot$, which is given in days.  The bottom table gives
the median activity, long-term and seasonal amplitudes, statistically
significant periods of variability ($P_\var$ in years), and the cycle
quality metric $Q_\cyc$.  The black bar to the left of the time series
visually shows the full range of measurements (short dashes), the
long-term amplitude $A_{98}$ (bar caps), short-term amplitude
$\overline{A}_{98,s}$ (thick bar), and the median activity
$\overline{S}$ (center diamond).  The highest three statistically
significant periodogram peaks are indicated by vertical green lines
along with their period.

Ensemble trends are still being analyzed, and final results will
appear in a future publication.  However, we will qualitatively
summarize some of our findings below.

\emph{Amplitude of variability scales with rotation and activity}.
Stars with faster rotation have larger amplitudes on both the
long-term and short-term time scales.  This was seen also in
\cite{Radick:1998} for long-term time scales using a sample of
FGK-type stars.  The increases in amplitude are significant.  Fast
rotating stars have about twice the solar cycle amplitude in
\emph{one year}.  There are linear trends in amplitude vs. median
activity.

\emph{Long cycles are found in the 50 year time series.} HD 20630 has
variability on two time scales, 5.7 years and 36 years, the latter
being remarkably long and only visible in these long time series.  A
single 38-year cycle is found in HD 224930 with a relatively high
$Q_\cyc$ = 44.  This cycle is easily identifiable and has a fast rise
and \emph{very} slow decay, similar to the solar cycle, but
exaggerated.  Addition of these long-term cycles to the famous
$P_\cyc$ vs $P_\rot$ plot of \cite{Bohm-Vitense:2007} introduces points
far above the two branches of activity discussed in that work,
complicating the discussion of multiple dynamo ``modes'' even
further.

\emph{Similar stars have similar patterns of long-term variability.}
Stellar twins identified by our distance metric appear to have similar
median activity levels and amplitudes of variability on long and short
time scales.  There are even indications that periods of variability
are shared among some close pairs.  This evidence seems to imply
stability in the stellar dynamo, which is not guaranteed given the
nonlinearity of the equations thought to govern the dynamo.

\emph{Very clear, clean cycles like the Sun are the minority.} Using
our cycle quality metric, only two other stars have cycles with
$Q_\cyc$ very close to the solar value of 59.  One of them, HD 81809,
has an $\sim$8 year cycle with $Q_\cyc = 61$, higher than the Sun.
However this signal is possibly due to a low-mass G9V component of the
binary \citep{Duquennoy:1988,Baliunas:1995}, which is mistakenly in
our sample due to its blended $(B-V)$ with an evolved, inactive
companion.  However, \citep{Pourbaix:2000} finds component masses of
1.7 and 1.0 solar masses, putting the low-mass component at the solar
value.  The properties of the source of this excellent cycle may only
be resolved by further spectroscopic observations able to separate the
components.  The second high-quality cycle comes from HD 197076, which
has an $\sim$5 year cycle with $Q_\cyc = 53$.  The luminosity,
temperature, and radius for this star are all equivalent to solar
within the measurement uncertainty, but no rotation
period is available.  The lower limit rotation period derived
from $v\sin(i)$ and the radius is $P_\rot > 18.7$.  Five more stars
with $Q_\cyc > 40$ have easily identifiable cycles that might be
subjectively classified as very ``solar like''.  Those stars (20 of
28) with $Q_\cyc < 40$ either have flat activity or tend to have more
erratic behavior that appears quite removed from the regularity of the
solar variations.  This is usually indicated by multiple significant
periods in the periodogram both above and below the ``main'' period of
variability, HD 30495 (Figure \ref{fig:hd30495}) being a good example
of this.

%%%
%%%
%%%
\section{Conclusion}

Diligent long-term observation programs by the Mount Wilson and Lowell
Observatory provide unique data for understanding the variability
patterns of Sun-like stars, with composite time series now approaching
50 years in length.  Questions on the uniqueness of the solar cycle,
and the sensitivity in stellar dynamos to changes in fundamental
properties can be approached using these data, improving our
understanding of the dynamo and our Sun in context.  Work is ongoing
to carefully quantify what these data can tell us about these
questions, but the initial results indicate that the clean, clear
solar cycle may be an exceptional case in the limited parameter regime
of solar analogs.

\bibliographystyle{cs19proc}
\bibliography{egeland_cs19}
%\bibliography{bibdesk-master}

\end{document}